\documentclass[amstex,12pt]{article}
\usepackage{amssymb}
\usepackage{amsmath}
\usepackage{IEEEtrantools}
\usepackage{pdflscape}
\usepackage{afterpage}
\usepackage{capt-of}

\usepackage{graphicx}
\usepackage{epstopdf}
\usepackage{subfig}

\usepackage{changepage}

\numberwithin{equation}{section}

\newcommand{\be}{\begin{equation}}
\newcommand{\ee}{\end{equation}}
\newcommand{\bea}{\begin{eqnarray}}
\newcommand{\eea}{\end{eqnarray}}
\newcommand{\non}{\nonumber}
\newcommand{\id}{\mathbb{I}}

\newcommand{\tr}{\mathop{\rm tr}\nolimits}

\usepackage[usenames,dvipsnames]{color}

\begin{document}

\begin{titlepage}
\strut\hfill UMTG--297
\vspace{.5in}
\begin{center}

\LARGE The spin-$s$ homogeneous central spin model:\\
exact spectrum and dynamics\\
\vspace{1in}
\large 
Rafael I. Nepomechie \footnote{Physics Department,
P.O. Box 248046, University of Miami, Coral Gables, FL 33124 USA, nepomechie@miami.edu}
and
Xi-Wen Guan \footnote{State Key Laboratory of Magnetic Resonance and Atomic and Molecular Physics,
Wuhan Institute of Physics and Mathematics, Chinese Academy of 
Sciences, Wuhan 430071, China, xwe105@wipm.ac.cn}
$\! {}^{, }$\footnote{Center for Cold Atom Physics, Chinese Academy of Sciences, Wuhan 430071, China}
$\! {}^{, }$\footnote{Department of Theoretical Physics, Research School of Physics and Engineering,
Australian National University, Canberra ACT 0200, Australia}
\\[0.8in]
\end{center}

\vspace{.5in}

\begin{abstract}
We consider the problem of a central spin with arbitrary spin $s$ that
interacts pairwise and uniformly with a bath of $N$ spins with
$s=1/2$.  We present two approaches for determining the exact spectrum
of this model, one based on properties of $SU(2)$, and the other based
on integrability.  We also analyze the exact time evolution of a spin
coherent state, and compute the time evolution of various quantities
of physical interest, including the entanglement entropy, spin
polarization and Loschmidt echo.
\end{abstract}

\end{titlepage}

\setcounter{footnote}{0}

\section{Introduction}\label{sec:intro}

The (spin-$\frac{1}{2})$ central spin model is a simple quantum mechanical
model of a central spin $\vec s_{0}$ interacting pairwise with a
``bath'' of $N$ surrounding spins $\vec s_{1}, \ldots, \vec s_{N}$, with
the Hamiltonian
\be
H^{(\frac{1}{2})}_{\rm{inhom}} = B s^{z}_{0} + 2 \sum_{j=1}^{N} A_{j} \vec s_{0} \cdot \vec s_{j} 
\,,
\ee
where $B$ and $A_{j}$ are real constants.
This is a special case of the 
Richardson-Gaudin model, which was formulated long ago 
\cite{Richardson:1963a, Richardson:1964a, Gaudin:1976sv, Gaudin:1983}. Nevertheless, 
this model is the focus of renewed attention due to its interesting 
new applications, such as quantum dots (see e.g. \cite{Khaetskii2002, Merkulov2002, Khaetskii2003, Erlingsson2004, 
Braun2005, Chen2007, Bortz2007, Bortz2010} and references therein). 
The homogeneous case $A_{j} = A$ is simple enough to allow for 
analytical analysis of spin dynamics, yet exhibits rich phenomena such as quantum collapse and 
revival \cite{Dooley2013, Guan:2018a, Guan:2018b}. Moreover, it can be mapped 
\cite{Dooley2013, Guan:2018b} to a generalization of the 
Jaynes-Cummings model \cite{Jaynes1963, Scully:1997, Bogoliubov2013} 
of quantum nonlinear optics.

We consider here a generalization of this model, whereby the central 
spin $\vec S_{0}$ has spin $s$. We focus primarily on the homogeneous 
case
\be
H^{(s)} = B S^{z}_{0} + 2 A \sum_{j=1}^{N} \vec S_{0} \cdot \vec 
s_{j}\,, \qquad s= \tfrac{1}{2}, 1, \ldots \,.
\label{spinShomogH}
\ee
This model, which to our knowledge 
has not been considered before, has the potential to be realized 
experimentally, and exhibits interesting quantum dynamics. It can 
also be mapped to a generalization of the Tavis-Cummings model 
\cite{Tavis1968, Scully:1997} of quantum nonlinear 
optics, see e.g. \cite{Knap2010, Youssef2010, Agarwal2012, Bogoliubov2013}
for recent work.

The outline of this paper is as follows.  In Section
\ref{sec:spectrum}, we discuss two approaches for determining the
exact spectrum of the model, one based on properties of $SU(2)$, and
the other based on integrability.  We also briefly describe the
connection with the Tavis-Cummings model.  In Section 3, we analyze the
exact time evolution of a spin coherent state.  Readers who are
primarily interested in spin dynamics can jump directly to this
section, as it is largely independent of the previous one. We 
conclude with a brief summary. Appendix \ref{sec:coeffs} contains the
derivation of a key step in our analysis of spin dynamics, while 
Appendix \ref{sec:identities} contains the derivation of some useful 
identities.

\section{Exact spectrum}\label{sec:spectrum}

We briefly discuss here two approaches for determining the spectrum of
the spin-$s$ homogeneous central spin model (\ref{spinShomogH}).  The
first approach exploits the partial $SU(2)$ symmetry of the problem,
and in principle can give the entire spectrum.  The second approach
exploits the integrability of the model; however, it remains to be understood 
whether the solution obtained in this way can give the full spectrum.

\subsection{$SU(2)$-based approach}\label{sec:su2}

The Hamiltonian (\ref{spinShomogH}) can evidently be rewritten as
\be
H^{(s)} = B S^{z}_{0} + 2 A  \vec S_{0} \cdot \vec J\,,
\label{spinShomogH2}
\ee
where
\be
\vec J = \sum_{j=1}^{N} \vec s_{j} 
\ee
is the total spin of the bath. This Hamiltonian clearly has the properties
\be
\left[ H^{(s)}\,, \vec{J}^{2} \right] = 0\,, \qquad \left[ H^{(s)}\,, \vec{S}_{0}^{2} \right] = 0\,, 
\label{properties}
\ee 
as well as the $U(1)$ symmetry
\be
\left[ H^{(s)}\,, {\cal S}^{z} \right] = 0\,, \qquad {\cal S}^{z} = J^{z} + 
S^{z}_{0}\,,
\label{U1symmetry}
\ee
but (for $B \ne 0$) does {\em not} have the full $SU(2)$ symmetry. In view of (\ref{properties}) and (\ref{U1symmetry}), we look for 
simultaneous eigenstates of $H$, ${\cal S}^{z}$, $\vec{S}_{0}^{2}$,  $\vec{J}^{2}$, 
\begin{align}
H^{(s)} |E, m, s, j \rangle &= E |E, m, s, j\rangle \,, \non \\
{\cal S}^{z} |E, m, s, j\rangle &= m |E, m, s, j\rangle \,, \non \\
\vec{S}_{0}^{2} |E, m, s, j\rangle &= s(s+1) |E, m, s, j\rangle \,, \non \\
\vec{J}^{2} |E, m, s, j\rangle &= j(j+1) |E, m, s, j\rangle \,.
\end{align}
We can expand these states in the standard orthonormal spin basis as follows
\be
|E, m, s, j\rangle = \sum_{m_{s}=-s}^{s}\sum_{m_{j}=-j}^{j} 
w_{m_{s}, m_{j}}^{(E,m,s,j)} \delta_{m_{s}+m_{j},m}|s, 
m_{s}\rangle\otimes |j, m_{j}\rangle\,,
\label{expansion}
\ee 
where the coefficients $w_{m_{s}, m_{j}}^{(E,m,s,j)}$ are still to be 
determined. Note that the allowed values of $j$ (spin of the bath) are
\be
j= \left\{
\begin{array}{ll}
    0, 1, \ldots, \frac{N}{2} & \mbox{ for } N = \mbox{ even } \\ \\
    \frac{1}{2},  \frac{3}{2}, \ldots, \frac{N}{2} & \mbox{ for } N = \mbox{ odd } 
    \end{array} \right. \,,
\label{jvalues}    
\ee
while the allowed values of $m$ are given by
\be
-j-s \le m \le j+ s \,.
\label{mvalues}  
\ee 
We now act on (\ref{expansion}) with the Hamiltonian (\ref{spinShomogH2}) in the form
\be
H^{(s)} = B S^{z}_{0} + A \left( S^{+}_{0} J^{-} +  S^{-}_{0} J^{+} + 2 S^{z}_{0} 
J^{z}\right) \,,
\label{spinShomogH3}
\ee
where $S^{\pm} = S^{x} \pm i S^{y}$ (and similarly for $J^{\pm}$). 
Using the familiar $SU(2)$ raising/lowering formulas 
\be
S^{\pm}| s, m\rangle 
= \sqrt{(s \mp m)(s \pm m +1)} | s, m\pm 1 \rangle \,,
\label{raisinglowering}
\ee
we obtain
\begin{align}
H^{(s)} |E, m, s, j\rangle &= \sideset{}{'}\sum_{m_{s}, m_{j}} w_{m_{s}, 
m_{j}}^{(E,m,s,j)} \Bigg\{
\mu_{m_{s}, m_{j}} |s, m_{s}\rangle\otimes |j, m_{j}\rangle \non\\
& + \nu_{m_{s}+1, m_{j}-1}^{(s,j)} |s, m_{s}+1\rangle\otimes |j, m_{j}-1\rangle 
+ \nu_{-m_{s}+1, -m_{j}-1}^{(s,j)}  |s, m_{s}-1\rangle\otimes |j, 
m_{j}+1\rangle \Bigg\}
\,,
\end{align}
where the summation is constrained by $m_{s}+m_{j}=m$, and the 
coefficients are defined by
\begin{align}
\mu_{m_{s}, m_{j}} &=  B\, m_{s} + 2 A\, m_{s}\, m_{j}  \,, \non \\
\nu_{m_{s}, m_{j}}^{(s,j)} &= A \sqrt{(s-m_{s}+1)(s+m_{s})(j+ 
m_{j}+1)(j-m_{j}) } \,.
\end{align}
Finally, using the orthonormality of the basis, we arrive at an 
eigenvalue relation for the energy $E$ and the corresponding
coefficients $w_{m_{s}, m_{j}}^{(E,m,s,j)}$,
\begin{align}
& \mu_{m_{s}, m_{j}}\, w_{m_{s}, m_{j}}^{(E,m,s,j)} +
\nu_{m_{s}, m_{j}}^{(s,j)} \, w_{m_{s}-1, m_{j}+1}^{(E,m,s,j)}  +
\nu_{-m_{s}, -m_{j}}^{(s,j)} \, w_{m_{s}+1, m_{j}-1}^{(E,m,s,j)} = E\, 
w_{m_{s}, m_{j}}^{(E,m,s,j)} \,, \non\\
& m_{s} = -s, \ldots, s\,, \qquad m_{j} = -j, \ldots, j\,, \qquad 
m_{s}+m_{j}=m \,.
\label{eigw}
\end{align}
Letting $d(m, s, j)$ denote the number of allowed values of $(m_{s}, m_{j})$ 
satisfying $m_{s}+m_{j}=m$, we see that (\ref{eigw}) entails 
diagonalizing a $d(m, s, j) \times d(m, s, j)$ matrix.

The eigenvalue relation (\ref{eigw}) for the energy is the main 
result of this subsection. Given values of $j$ and $m$ (constrained 
by (\ref{jvalues}) and  (\ref{mvalues}), respectively), the 
eigenvalue problem (\ref{eigw}) can in principle be solved for the 
corresponding energies. The advantage of this approach over 
brute-force diagonalization of the full Hamiltonian (\ref{spinShomogH2})
is that the matrices to be diagonalized are much smaller. Of course, as 
the bath size $N$ becomes large, the number of possible values for $j$ and $m$ 
also becomes large.

As a simple example, we present in Table \ref{table:s1N2} the energies
that are computed in this way for the case $s=1\,, N=2$.  (The final
two columns of the table are explained in the following subsection.)

\begin{table}[htb]
    \small 
  \centering
  \begin{tabular}{|c|c|c|c|c|}
     \hline
      $j$ & $m$ & $E$ & $M$ &  $\{ v_{a} \}$\\
    \hline
     1 & 2 & 1.5 & 0 & -\\
     1 & 1 & -0.780776 & 1 & -0.438447\\
     1 & 1 & 1.28078   & 1 &  -4.56155\\
     1 & 0 & -2.14854 & 2 & -0.351465 $\pm 0.262932 i$ \\
     1 & 0 & -0.893401 & 2 & -2.71954, -0.493659 \\
     1 & 0 &  1.04194 & 2 & -3.54194 $\pm  1.70866 i$ \\
     1 & -1 & -1.28078 & 3 & -0.612504, -1.41297 $\pm 0.681796 i$\\
     1 & -1 & 0.780776 & 3 & -3.16744, -2.19705 $\pm 2.46224 i$  \\
     1 & -2 & 0.5 & 4 & -2.26566 $\pm 0.850941 i$, -0.734342 $\pm 
     2.43893 i$\\
     0 & 1 &  0.5 & & \\
     0 & 0 & 0  & & \\
     0 & -1 & - 0.5 & & \\
        \hline
   \end{tabular}
   \caption{\small The energies ($E$) and Bethe roots $\{ v_{1}, 
   \ldots, v_{M} \}$ for the Hamiltonian (\ref{spinShomogH}) with 
   $s=1, N=2, A = B = 0.5$. The total number of levels is $(2s+1) 2^{N} = 12$.}
   \label{table:s1N2}
\end{table}

\subsection{Bethe ansatz approach}

The spin-$s$ central spin model (\ref{spinShomogH}) is integrable, as is the 
original model with $s=\frac{1}{2}$. We present here its Bethe ansatz 
solution, but only sketch the derivation, since it is similar to the one 
for the spin-$\frac{1}{2}$ case, see e.g. \cite{Sklyanin:1987ih, 
Hikami:1992np, Zhou:2001}. For pedagogical reasons,
we in fact work for the inhomogeneous model (which is also integrable), and consider the 
homogeneous limit only at the end.

\subsubsection{Inhomogeneous case}

Our starting point is the $SU(2)$-invariant $(\frac{1}{2},s)$ R-matrix on ${\cal C}^{2} 
\otimes {\cal C}^{2s+1}$ (see, e.g. 
\cite{Babujian:1983ae}),
\be
R^{(\frac{1}{2},s)}(u)=\frac{1}{u+(s+\frac{1}{2})\eta}
\left((u+\frac{\eta}{2})\id + 2\eta \vec s \overset{\cdot}{\otimes} 
\vec S \right) \,, \qquad s = \tfrac{1}{2}\,, 1\,, \ldots\,,
\ee
where (as above) $\vec s$ and  $\vec S$ denote the spin operators for 
spin-$\frac{1}{2}$ and spin-$s$, respectively. This R-matrix 
is a solution of the Yang-Baxter equation on ${\cal C}^{2} 
\otimes {\cal C}^{2} \otimes {\cal C}^{2s+1}$
\be
R^{(\frac{1}{2}, \frac{1}{2})}_{12}(u-v)\, R^{(\frac{1}{2}, s)}_{13}(u)\, R^{(\frac{1}{2}, s)}_{23}(v) = 
R^{(\frac{1}{2}, s)}_{23}(v)\, R^{(\frac{1}{2}, s)}_{13}(u)\, 
R^{(\frac{1}{2}, \frac{1}{2})}_{12}(u-v) \,.
\label{YBE}
\ee
We introduce the following monodromy matrix
\be
T_{a}(u) = G_{a}\, R^{(\frac{1}{2},\frac{1}{2})}_{a 
N}(u-\epsilon_{N})\, \ldots R^{(\frac{1}{2},\frac{1}{2})}_{a 
1}(u-\epsilon_{1})\, R^{(\frac{1}{2},s)}_{a 0}(u-\epsilon_{0})\,, 
\qquad G = e^{\eta s B \sigma^{z}} \,,
\ee
where the ``auxiliary'' space (denoted by $a$) has dimension 2, and there 
are $N+1$ ``quantum'' spaces (denoted by $0, 1, \ldots, N$): the 
$0^{th}$ quantum space (corresponding to the central spin) has 
dimension $2s+1$, while all the others  (corresponding to the bath) 
have dimension 2. Notice that there are arbitrary inhomogeneities 
$\epsilon_{0}, \ldots, \epsilon_{N}$ associated with each of the 
quantum spaces. Finally, note that there is a diagonal twist, encoded 
by the matrix $G$, which is responsible for breaking $SU(2)$ down to 
$U(1)$.

The transfer matrix $t(u)$, obtained by tracing the monodromy matrix 
over the auxiliary space
\be
t(u) = \tr_{a} T_{a}(u) \,,
\label{transfer}
\ee
satisfies the important commutativity property
\be
\left[ t(u) \,, t(v) \right] = 0
\ee
owing to the Yang-Baxter equation (\ref{YBE}). 

The basic idea, following e.g. \cite{Sklyanin:1987ih, Hikami:1992np, Zhou:2001},
is to evaluate the transfer matrix at $u=\epsilon_{0}$ and expand in terms 
of $\eta$. We find that the inhomogeneous spin-$s$ Hamiltonian
\be
H^{(s)}_{\rm{inhom}} = B S^{z}_{0} +  \frac{1}{s}\sum_{j=1}^{N} 
\frac{1}{\epsilon_{0}-\epsilon_{j}} \vec S_{0} \cdot \vec s_{j} 
\label{spinSinhomogH}
\ee
can indeed be obtained in this way
\be
H^{(s)}_{\rm{inhom}} = \frac{(2s+1)}{4s} 
\frac{d}{d\eta}t(\epsilon_{0})\Big\vert_{\eta=0} + 
\frac{1}{4s}\sum_{j=1}^{N}\frac{1}{\epsilon_{0}-\epsilon_{j}} \id \,.
\label{Hfromtransf}
\ee
Using algebraic Bethe ansatz (ABA), we find that the eigenvalues of the
transfer matrix (\ref{transfer}) are given by
\be
\Lambda(u) = e^{\eta s B} \prod_{a=1}^{M} 
\frac{u-v_{a}-\eta}{u-v_{a}} +
e^{-\eta s B}  
\left(\frac{u-\epsilon_{0}+(\frac{1}{2}-s)\eta}
{u-\epsilon_{0}+(\frac{1}{2}+s)\eta}\right)
\prod_{j=1}^{N}\frac{u-\epsilon_{j}}{u-\epsilon_{j}+\eta}
\prod_{a=1}^{M} 
\frac{u-v_{a}+\eta}{u-v_{a}} \,,
\ee
where $\{v_{1}\,, \ldots\,, v_{M} \}$ are solutions of the Bethe equations
\be
\left(\frac{v_{a}-\epsilon_{0}+(\frac{1}{2}-s)\eta}
{v_{a}-\epsilon_{0}+(\frac{1}{2}+s)\eta}\right)
\prod_{j=1}^{N}\frac{v_{a}-\epsilon_{j}}{v_{a}-\epsilon_{j}+\eta}
= e^{2\eta s B} \prod_{b=1, b\ne a}^{M} 
\frac{v_{a}-v_{b}-\eta}{v_{a}-v_{b}+\eta}\,, \quad a = 1, \ldots, M
\,.
\label{BA1}
\ee
It follows from (\ref{Hfromtransf}) that the eigenvalues of 
$H^{(s)}_{\rm{inhom}} $ are given by
\begin{align}
E_{\rm{inhom}} &=  \frac{(2s+1)}{4s} 
\frac{d}{d\eta}\Lambda(\epsilon_{0})\Big\vert_{\eta=0} + 
\frac{1}{4s}\sum_{j=1}^{N}\frac{1}{\epsilon_{0}-\epsilon_{j}} \non \\
& = s B  + 
\frac{1}{2}\sum_{j=1}^{N}\frac{1}{\epsilon_{0}-\epsilon_{j}} + 
\sum_{a=1}^{M}\frac{1}{v_{a}-\epsilon_{0}}\,.
\label{energyinhom}
\end{align}
Moreover, expanding the Bethe equations (\ref{BA1}) in $\eta$, we 
arrive at the Bethe equations for the inhomogeneous spin-$s$ central spin model
\be
-2 s B  -  \frac{2s}{v_{a} - \epsilon_{0}}
- \sum_{j=1}^{N} \frac{1}{v_{a} - \epsilon_{j}} 
+ 2 \sum_{b=1, b\ne a}^{M} \frac{1}{v_{a}-v_{b}} = 0\,, \quad a = 1, \ldots, M \,.
\label{BAinhom}
\ee 
The allowed values of $M$ can be deduced from the formula for the  
${\cal S}^{z}$ eigenvalues
\be
m = \frac{N}{2} + s - M
\ee
which also follows from the ABA, together with the ranges 
(\ref{jvalues}) and (\ref{mvalues}). We conclude that the number of 
Bethe roots can be
\be
M = 0, 1, \ldots, N+2s\,.
\label{Mvalues}
\ee 

The energy formula (\ref{energyinhom}) and the corresponding Bethe 
equations (\ref{BAinhom}) constitute our main results for the 
inhomogeneous spin-$s$ central spin model (\ref{spinSinhomogH}). For 
$s=\frac{1}{2}$, the well-known results are recovered.
For distinct values of the inhomogeneities ($\epsilon_{j} \ne 
\epsilon_{k}$ for $j \ne k$) and $B \ne 0$, the spectrum is 
nondegenerate, and the Bethe ansatz solution appears to be complete. 
(We have checked this numerically for small values of $s$ and $N$; a proof 
for the case $s=\frac{1}{2}$ can be found in \cite{Links:2017}.) We 
conjecture that the number of solutions of these Bethe equations for 
given values of $s, N, M$ is given by
\be
{\cal N}(s, N, M) = \sum_{k=0}^{\lfloor s \rfloor} (-1)^{k} 
{2s-k\choose k} {N+2s-2k\choose m-k}\,.
\ee
We have also checked this result numerically for small values of $s$ and $N$, 
and one can verify that indeed
\be
\sum_{M=0}^{N+2s} {\cal N}(s, N, M) = (2s+1) 2^{N} \,,
\ee
thereby accounting for all the levels of the system.

\subsubsection{Homogeneous case}

Let us finally return to the homogeneous Hamiltonian 
(\ref{spinShomogH}). Comparing with the inhomogeneous one 
(\ref{spinSinhomogH}), we see that the homogeneous case corresponds 
to setting
\be
\epsilon_{1} = \ldots = \epsilon_{N} \equiv \epsilon\,, \qquad 
A = \frac{1}{2s(\epsilon_{0} - \epsilon)} \,.
\ee 
For these values of parameters, the formulas for the energy (\ref{energyinhom}) and the Bethe 
equations (\ref{BAinhom}) reduce to 
\be
E = s (B + N A)  + \sum_{a=1}^{M}\frac{1}{v_{a}}
\label{EBAhomg}
\ee
and
\be
-2 s B  -  \frac{2s}{v_{a}}
- \frac{N}{v_{a} + \frac{1}{2 s A}} 
+ 2 \sum_{b=1, b\ne a}^{M} \frac{1}{v_{a}-v_{b}} = 0\,, \quad a = 1, \ldots, M \,,
\label{BAhom}
\ee 
respectively, after performing a shift $v_{a} \mapsto v_{a} 
+  \epsilon_{0}$ of all the Bethe roots. Eqs. (\ref{EBAhomg}) and 
(\ref{BAhom}) constitute our main results for the Bethe ansatz 
solution of the homogeneous model.

A simple example with $s=1$ and $N=2$ is presented in Table  \ref{table:s1N2}. Note that
all the levels with $j = \frac{N}{2}$ are accounted for, but
not those with $j < \frac{N}{2}$. This appears to be a general 
feature of the Bethe ansatz solution for the homogeneous 
model. Our preliminary investigations indicate that this
difficulty (which is present already for the spin-$\frac{1}{2}$ case)
is due to the necessity of correctly taking into account so-called singular solutions 
(e.g., $v=0$).\footnote{For a recent discussion of singular solutions in 
the context of the periodic Heisenberg chain, see \cite{Hao:2013jqa}.} We hope 
to investigate this matter further in the future.

\begin{figure}[htb]
\centering
{\includegraphics[width=5.5cm]{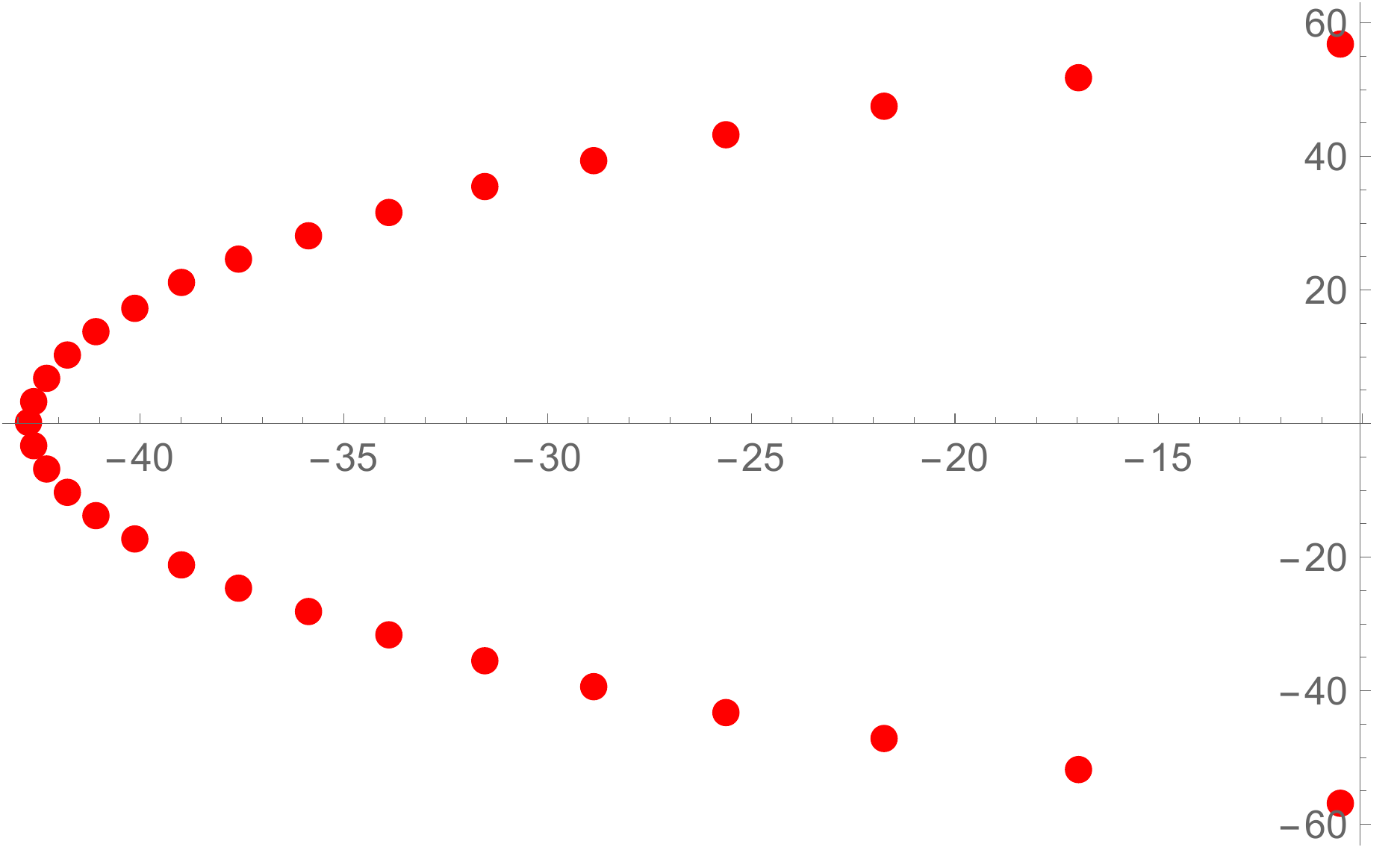}}
\caption{A set of solutions of the Bethe equations (\ref{BAhom}) 
plotted in the complex $v$ plane for 
$N=60, s=1, M=31, A=B=0.5$, with corresponding energy $E=30.004$.} 
\label{fig:BetherootsN60M31}
\end{figure}

A nice feature of the Bethe equations (\ref{BAhom}) is that numerical solutions 
can be readily found even for relatively large values of $N$ and $M$. 
The trick (see e.g. \cite{Gaudin:1971zza, Shastry:2001, Pan:2011, Marquette:2012} 
begins with the observation that the q-polynomial 
\be
q(u) = \prod_{a=1}^{M}(u-v_{a}) = u^{M} + O(u^{M-1}) 
\ee
satisfies
\be
\frac{q''(v_{a})}{q'(v_{a})} = 2 \sum_{b=1, b\ne a}^{M} 
\frac{1}{v_{a}-v_{b}} \,.
\ee
The Bethe equations (\ref{BAhom}) can therefore be rewritten in the 
form 
\be
P(v_{a}) = 0\,, \qquad a = 1, \ldots, M \,,
\ee 
where
\be
P(u) = u (u + \frac{1}{2 s A}) q''(u)
- 2 s B u (u + \frac{1}{2 s A}) q'(u)  - 2 s (u + \frac{1}{2 s A}) q'(u) - N u q'(u) \,.
\ee 
One next observes that $P(u)$ is a polynomial in $u$ of degree $M+1$, which 
has all $M$ zeros of the polynomial $q(u)$. Therefore, $P(u)/q(u)$ is 
a polynomial of degree 1, i.e.
\be
P(u) = (a + b\, u)\, q(u) \,,
\label{Pq}
\ee
where $b= - 2 s B M$ follows from the asymptotic behavior $u 
\rightarrow \infty$. Setting
\be
q(u) = \sum_{k=0}^{M} q_{k} u^{k} \,,
\ee
one can obtain from (\ref{Pq}) (by setting the coefficients of 
$u^{0}\,, \ldots\,, u^{M+1}$ equal to zero)
a set of $M+2$ equations for the  $M+2$ unknowns $a, 
q_{0}, \ldots, q_{M}$, which can be readily solved numerically even for 
relatively large values of $N$ and $M$; one can then
determine the zeros of $q(u)$, which are the sought-after Bethe 
roots.\footnote{Unfortunately, this trick is not nearly as effective 
for the inhomogeneous case.}
An example with $N=60, s=1, M=31$ is shown in Figure 
\ref{fig:BetherootsN60M31}.

\subsection{Connection with the Tavis-Cummings model}

We briefly note here a mapping of the spin-$s$ homogeneous central 
spin model to the Tavis-Cummings model \cite{Tavis1968, Scully:1997}.
This mapping relies, as in the spin-$\frac{1}{2}$ 
case \cite{Dooley2013, Guan:2018b}, on the Holstein-Primakoff 
transformation \cite{Holstein1940}:
\be
J^{+} = \sqrt{N} a^{\dagger} \sqrt{1-\frac{a^{\dagger} a}{N}} \,, 
\qquad J^{-} = \left( J^{+} \right)^{\dagger} 
=\sqrt{N}\sqrt{1-\frac{a^{\dagger} a}{N}}\, a \,, \qquad 
J^{z} = a^{\dagger}  a - \frac{N}{2} \,,
\label{HP}
\ee
where $\left[ a \,, a^{\dagger} \right] = 1$. Indeed, let us now consider 
an anisotropic generalization of the Hamiltonian (\ref{spinShomogH3}), 
\be
H^{(s)} = B S^{z}_{0} + A \left( S^{+}_{0} J^{-} +  S^{-}_{0} J^{+} 
\right) + 2 \Delta\, S^{z}_{0} J^{z} \,,
\ee
where $\Delta$ is the anisotropy parameter.\footnote{For $\Delta=A$, this 
model evidently reduces to the isotropic model (\ref{spinShomogH3}). For simplicity, we focus in this paper 
primarily on the isotropic case; however, it is possible to generalize 
all of the results to the anisotropic case.} Applying the  
transformation (\ref{HP}), and letting $N \rightarrow \infty$, we obtain
\be
H^{(s)} \sim (B- \Delta N) S^{z}_{0} + A \sqrt{N} \left(S^{+}_{0}\, a + 
S^{-}_{0}\, a^{\dagger} \right) + 2 \Delta S^{z}_{0}\, a^{\dagger} a 
\,.
\ee
The case $\Delta = 0$ reduces to the Tavis-Cummings model, while the 
isotropic case $\Delta=A$ corresponds to a generalization of the Tavis-Cummings model.

\section{Exact dynamics}\label{sec:dynamics}

We turn now to the question of how states evolve in time
for the spin-$s$ homogeneous central spin model (\ref{spinShomogH}),  
(\ref{spinShomogH2}). For 
a generic initial state, this problem appears to pose a formidable challenge, 
even for a model as simple as this one. However, if the initial 
state is an eigenstate of $\vec{S}_{0}^{2}$ and 
$\vec{J}^{2}$, then the symmetries (\ref{properties}) and 
(\ref{U1symmetry}) significantly constrain the possible intermediate 
states, and the problem becomes tractable yet nevertheless remains nontrivial.

\subsection{Time evolution of a spin coherent state}

Following \cite{Dooley2013, Guan:2018a}, we assume that the bath is 
initially in a so-called spin coherent state \cite{Radcliffe1971, Arecchi1972}
\be
|\theta \rangle = \bigotimes_{j=1}^{N}\left[\cos(\tfrac{\theta}{2}) 
|\tfrac{1}{2}, \tfrac{1}{2}\rangle_{j} + \sin(\tfrac{\theta}{2})  
|\tfrac{1}{2}, -\tfrac{1}{2}\rangle_{j} \right]\,,
\label{spincoherentstate}
\ee
which indeed is an eigenstate of $\vec{J}^{2}$ with $j=\frac{N}{2}$. 
Moreover, we assume that the central spin is initially ``up'', i.e. 
in the state $|s, s \rangle_{0}$. Thus, the initial state of the 
system is
\be
|\Psi(0)\rangle = |s, s \rangle \otimes |\theta \rangle \,,
\label{initialstate}
\ee
and our task is to determine its time evolution
\be
|\Psi(t)\rangle = e^{-i H^{(s)} t} |\Psi(0)\rangle \,.
\ee

Expressing the spin coherent state in terms of so-called Dicke states 
as in \cite{Dooley2013, Guan:2018b}
\be
|\theta \rangle = \sum_{n=0}^{N} \sqrt{{N \choose n}} 
\cos^{N-n}(\tfrac{\theta}{2})  \sin^{n}(\tfrac{\theta}{2})  
|n\rangle\,,
\ee
where $|n\rangle \equiv |\frac{N}{2},\frac{N}{2} -n \rangle$, the 
problem reduces to computing
\begin{align}
|\Psi(t)\rangle &=   \sum_{k=0}^{\infty} \frac{1}{k!} (-i H^{(s)} t)^{k} 
|\Psi(0)\rangle \non \\
&= \sum_{n=0}^{N} \sqrt{{N \choose n}} 
\cos^{N-n}(\tfrac{\theta}{2})  \sin^{n}(\tfrac{\theta}{2})  
\sum_{k=0}^{\infty} \frac{(-i t)^{k}}{k!}  (H^{(s)})^{k} \left(|s, s \rangle 
\otimes |n\rangle \right) \,.
\label{intermed}
\end{align}
Due to the $U(1)$ symmetry (\ref{U1symmetry}), we know that
\be
(H^{(s)})^{k} \left(|s, s \rangle \otimes |n\rangle \right) = 
\sum_{j=0}^{2s} h^{(s, k)}_{j} 
|s,s-j \rangle \otimes |n-j\rangle \,,
\label{Hk}
\ee
where the coefficients $h^{(s, k)}_{j}$ are still unknown. 
We show in Appendix \ref{sec:coeffs} that these coefficients are given by
\be
h_{j}^{(s,k)} = \sum_{l=0}^{2s} c_{j,l}^{(s)}(n)\, 
\left(\omega_{l}^{(s)}(n)\right)^{k}  \,, \qquad j = 0, 1, \ldots, 2s 
\,,
\label{hjsltnmain}
\ee 
see (\ref{hjsltn}), and we provide a straightforward recipe for 
numerically computing $c_{j,l}^{(s)}(n)$ and $\omega_{l}^{(s)}(n)$.
Substituting the results (\ref{Hk}) and (\ref{hjsltnmain}) back into 
(\ref{intermed}), we obtain
\begin{align}
|\Psi(t)\rangle &=  \sum_{n=0}^{N} \sqrt{{N \choose n}} 
\cos^{N-n}(\tfrac{\theta}{2})  \sin^{n}(\tfrac{\theta}{2})  
\sum_{j=0}^{2s}
\sum_{l=0}^{2s} c_{j,l}^{(s)}(n)\, 
\sum_{k=0}^{\infty} \frac{(-i t)^{k}}{k!} 
\left(\omega_{l}^{(s)}(n)\right)^{k}
|s,s-j \rangle \otimes |n-j\rangle \non \\
&= \sum_{n=0}^{N} \sqrt{{N \choose n}} 
\cos^{N-n}(\tfrac{\theta}{2})  \sin^{n}(\tfrac{\theta}{2})  
\sum_{j, l =0}^{2s}
c_{j,l}^{(s)}(n)\, 
e^{-i t \omega_{l}^{(s)}(n)}
|s,s-j \rangle \otimes |n-j\rangle \,.
\label{evolvedstate}
\end{align}
For given values of $s$, $N$ and $n$, the
frequencies $\{\omega_{0}^{(s)}(n)\,, \ldots\,, 
\omega_{2s}^{(s)}(n) \}$ are the energies in the sector (see Sec. \ref{sec:su2})
with $j=\frac{N}{2}$ and $m = \frac{N}{2} - n + s$, which has (at 
most) dimension $2s+1$.

An important check on this result is the verification of unitarity $\langle 
\Psi(t) | \Psi(t)  \rangle = 1$. Using the 
orthonormality of the basis and the fact that 
$c_{j,l}^{(s)}(n)$ and $\omega_{l}^{(s)}(n)$ are real, we obtain
\be
\langle \Psi(t) | \Psi(t)  \rangle =
\sum_{n=0}^{N} {N \choose n}
\cos^{2(N-n)}(\tfrac{\theta}{2})  \sin^{2n}(\tfrac{\theta}{2})  
\sum_{j, l, l' =0}^{2s}
c_{j,l}^{(s)}(n)\, c_{j,l'}^{(s)}(n)\, 
e^{-i t \left(\omega_{l}^{(s)}(n) - \omega_{l'}^{(s)}(n) \right)} \,.
\ee
With the help of the identities (see (\ref{id1}) and (\ref{id5})) 
\be
\sum_{j=0}^{2s} c_{j,l}^{(s)}(n) \, c_{j,l'}^{(s)}(n)  = 
c_{0,l}^{(s)}(n)\, 
\delta_{l, l'}\,, \qquad \sum_{l=0}^{2s} c_{0,l}^{(s)}(n) = 1\,,
\label{identities}
\ee
we see that unitarity is indeed preserved.

The result (\ref{evolvedstate}) is one of the main results of our 
paper. We emphasize that this is an exact result. Note that the infinite sum over $k$ in (\ref{intermed}) has 
been effectuated, leaving only finite sums to be performed (assuming 
that $N$ and $s$ are finite). For the special case 
$s=\frac{1}{2}$, the result (\ref{evolvedstate}) reduces to a corresponding result in 
\cite{Guan:2018b}. In the remaining part of this section, we use  
(\ref{evolvedstate}) to compute the time evolution of various quantities of physical interest.

\subsection{Reduced density matrix}

The reduced density matrix for the central spin $\rho(t)$ is defined by
\be
\rho(t) = \sum_{\alpha} \langle \alpha| \Psi(t)\rangle \langle \Psi(t) | 
\alpha \rangle \,,
\ee
where the trace is performed by summing over an orthonormal basis of 
the bath $({\cal C}^{2})^{\otimes N}$. Making use of the result 
(\ref{evolvedstate}) and the fact
\be
\sum_{\alpha} \langle \alpha| n-j\rangle \langle n'-j' | 
\alpha \rangle = \delta_{n-j,n'-j' } \,,
\ee
we obtain the $(2s+1) \times (2s+1)$ matrix
\be
\rho(t) = \sum_{j, j' =0}^{2s} \rho_{j j'}(t) |s-j\rangle \langle s-j'| 
\,,
\ee
whose matrix elements are given by
\begin{align}
\rho_{j j'}(t)  &= \sum_{n, n'=0}^{N} 
\sum_{l, l' =0}^{2s}
\sqrt{{N \choose n} {N \choose n'}} 
\cos^{2N-n-n'}(\tfrac{\theta}{2})  \sin^{n+n'}(\tfrac{\theta}{2})\,  
\delta_{n-j,n'-j'} \non \\
& \qquad \qquad 
\times c_{j,l}^{(s)}(n)\, c_{j',l'}^{(s)}(n')\, e^{-i t 
\left(\omega_{l}^{(s)}(n) - \omega_{l'}^{(s)}(n') \right)} \,.
\label{rhomat}
\end{align}
The matrix (\ref{rhomat}) is manifestly Hermitian, $\rho^{\dagger}=\rho$.

\begin{figure}[htb]
\centering
{\includegraphics[width=5.5cm]{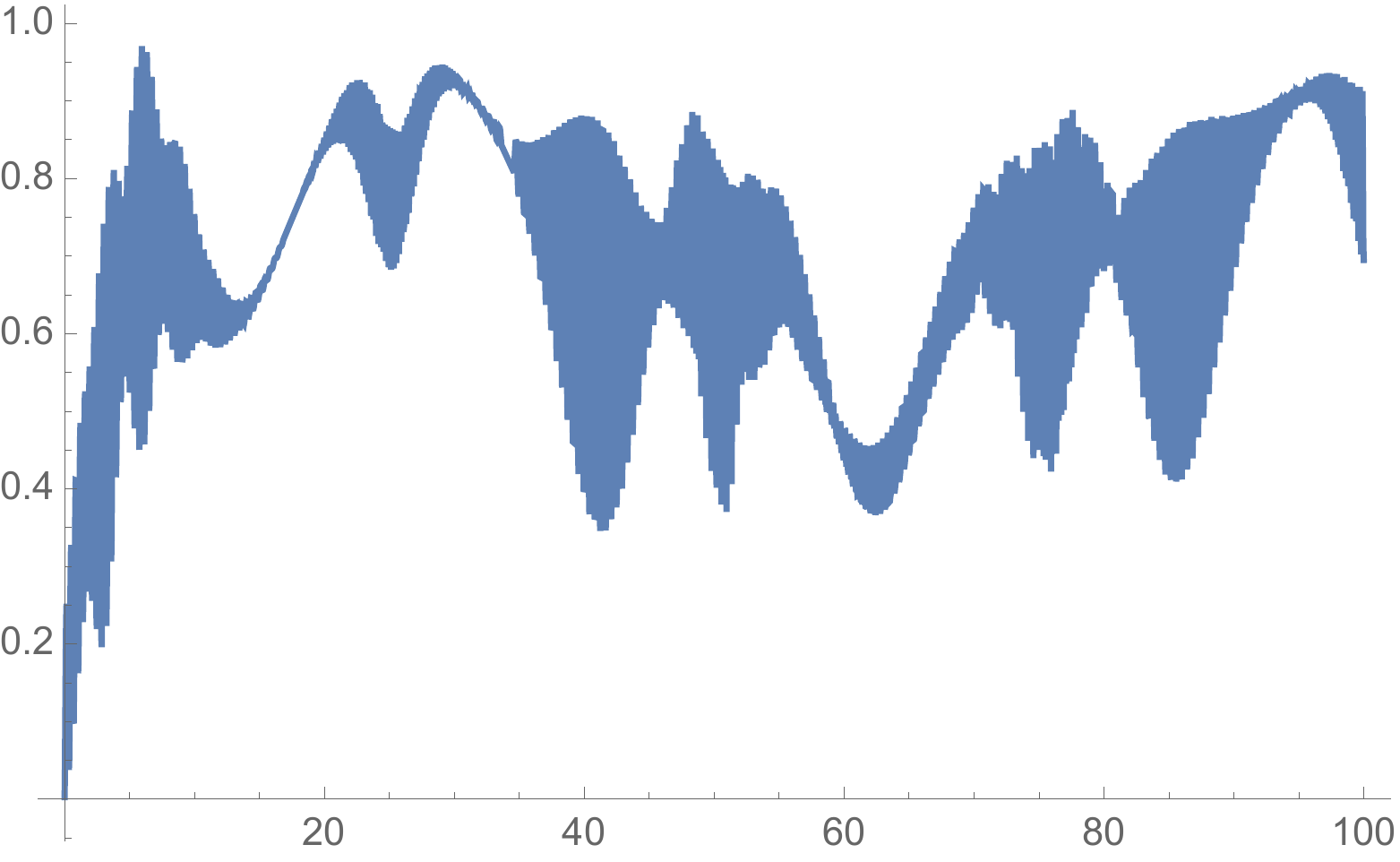}}
\caption{The von Neumann entanglement entropy (\ref{entropy})
as a function of time for $N=15, s=1, \theta=0.5 \pi, A=B=1.0$.} 
\label{fig:entanglement_entropy}
\end{figure}

Knowing the reduced density matrix, we can directly compute
the von Neumann entanglement entropy $S(t)$
\be
S(t) = - \tr \left[ \rho(t) \ln \rho(t) \right] = - \sum_{l=0}^{2s} 
\lambda_{l}(t) \ln \lambda_{l}(t) \,,
\label{entropy}
\ee
and the quantum purity $\gamma(t)$
\be
\gamma(t) = \tr \left[ \rho^{2}(t) \right] = \sum_{l=0}^{2s} 
\lambda_{l}^{2}(t) \,,
\ee
where $\{ \lambda_{l}(t) \}$ are the eigenvalues of $\rho(t)$.
An example of the von Neumann entanglement entropy $S(t)$ for $s=1$ is presented in 
Fig. \ref{fig:entanglement_entropy}. In contrast with the 
$s=\frac{1}{2}$ case \cite{Guan:2018b}, here $S(t)$ displays rapid 
irregular oscillations except at the collapsed regions. (See also 
Fig. \ref{fig:spin_polarization}.)

\subsection{Spin expectation value}

\begin{figure}[htb]
\centering
{\includegraphics[width=5.5cm]{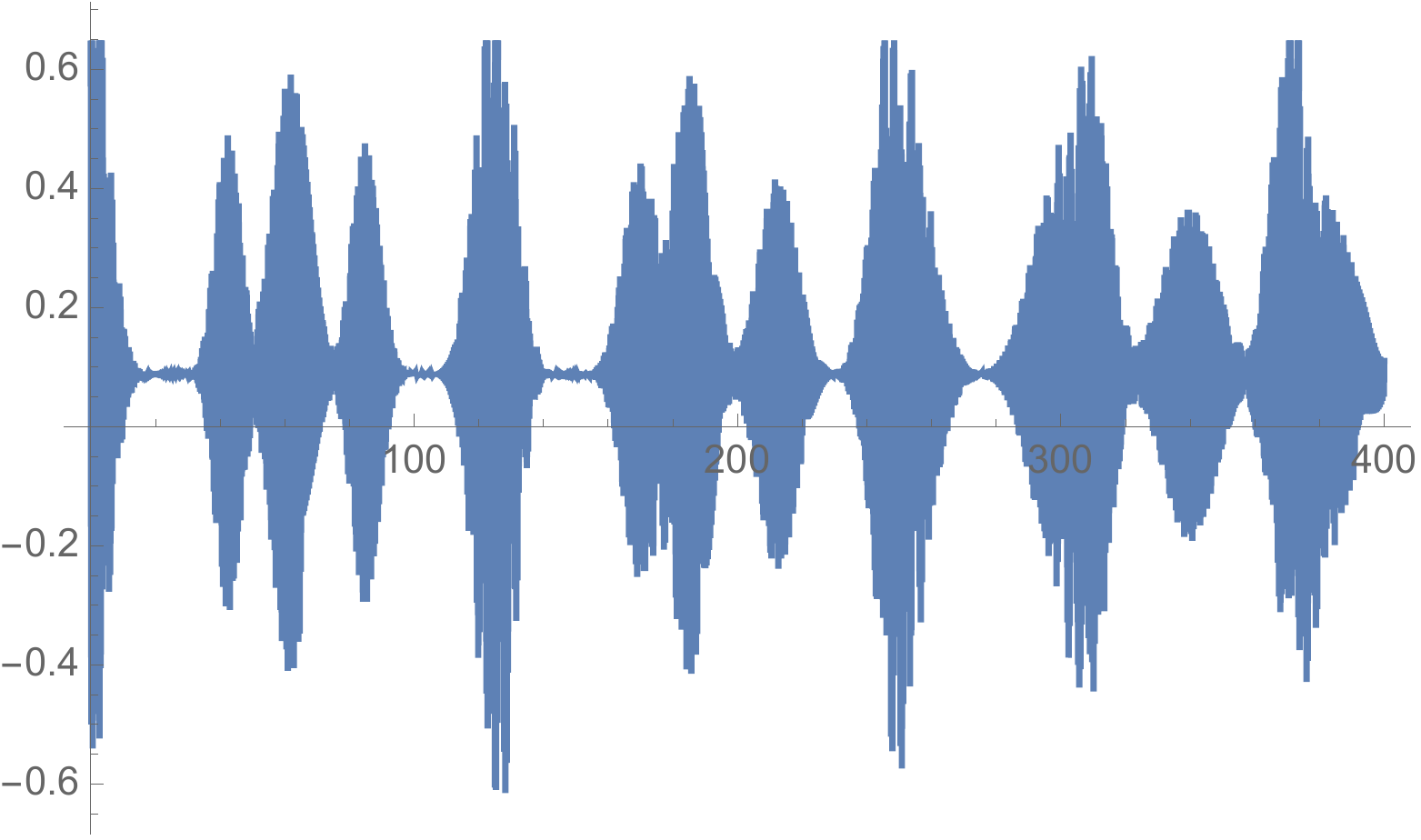}}
\caption{The spin polarization of the central spin (\ref{spinpolarization})
as a function of time for $N=15, s=1, \theta=0.5 \pi, A=B=1.0$.} 
\label{fig:spin_polarization}
\end{figure}

The expectation value of the central spin can be computed using the 
reduced density matrix
\be
\vec S(t) = \langle \Psi(t) | \vec S | \Psi(t) \rangle = \tr \left[ 
\vec S\, \rho(t) \right] \,.
\ee
In particular, the so-called spin polarization is given in terms of the 
diagonal elements of (\ref{rhomat})
\be
S^{z}(t) = \sum_{n=0}^{N} 
\sum_{j, l, l' =0}^{2s}
{N \choose n} 
\cos^{2(N-n)}(\tfrac{\theta}{2})  \sin^{2n}(\tfrac{\theta}{2})\,  
(s-j)\, c_{j,l}^{(s)}(n)\, c_{j,l'}^{(s)}(n)\, e^{-i t 
\left(\omega_{l}^{(s)}(n) - \omega_{l'}^{(s)}(n) \right)} \,.
\label{spinpolarization}
\ee 
An example of the spin polarization for $s=1$ is presented in 
Fig. \ref{fig:spin_polarization}. In contrast with the 
$s=\frac{1}{2}$ case \cite{Guan:2018b}, the duration of collapses 
quickly tend to zero as $t$ increases. This indicates 
that the bath is mixing all the states of the central spin. Note also 
that, for spin $s$, the revival regions consist of up to $2s+1$ 
revival peaks.

\begin{figure}[htb]
\centering
{\includegraphics[width=5.5cm]{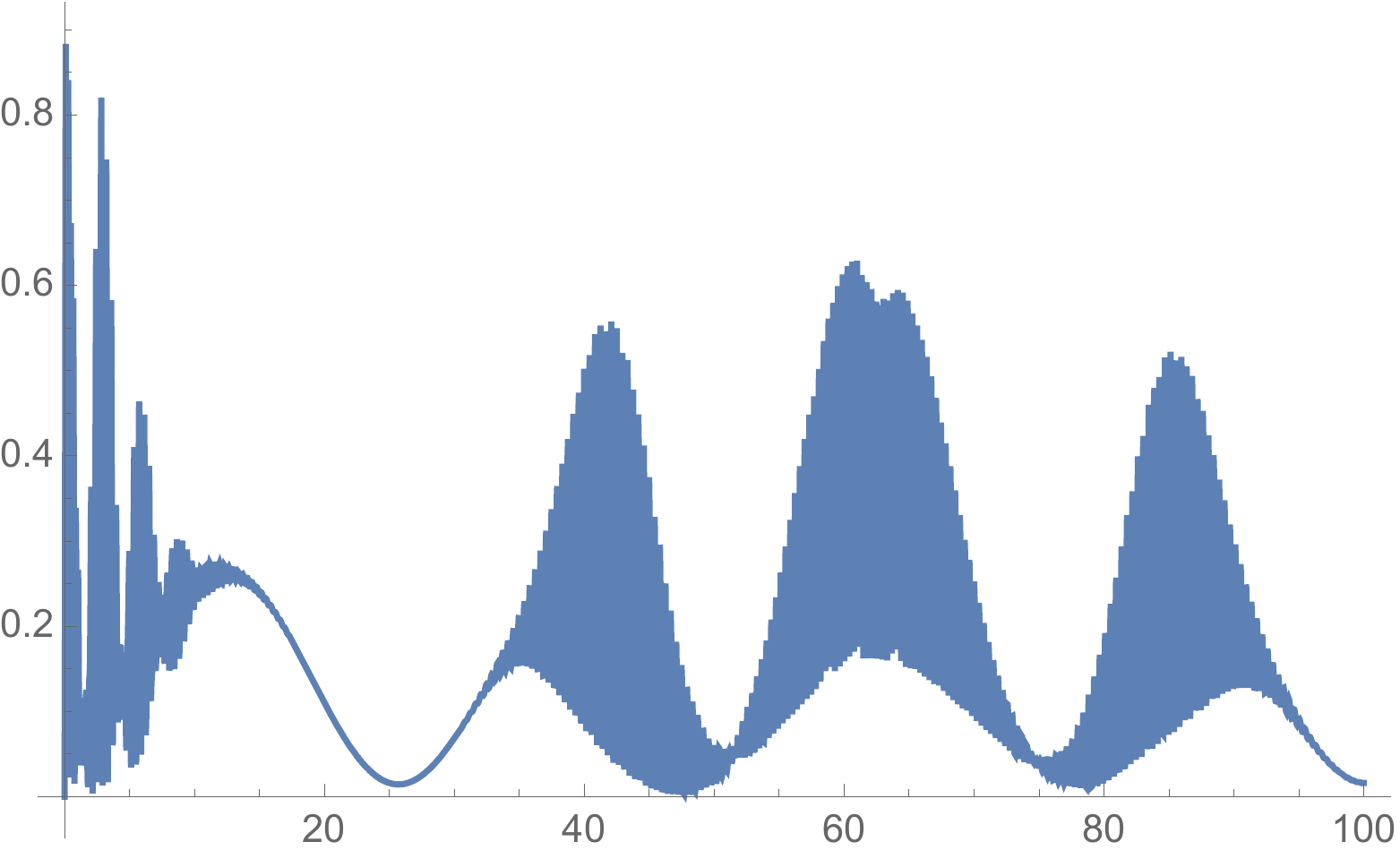}}
\caption{The squared norm coherent factor 
of the central spin $|S^{-}(t)|^{2}$ as a function of time for $N=15, 
s=1, \theta=0.5 \pi, A=B=1.0$.} 
\label{fig:coherentfactor}
\end{figure}

On the other hand, the so-called coherent factor $S^{-}(t) = 
\langle \Psi(t) | S^{-} | \Psi(t) \rangle$ involves 
off-diagonal elements of the reduced density matrix, an example of 
which is presented in Fig. \ref{fig:coherentfactor}.

\subsection{Loschmidt echo}

The Loschmidt echo can also be readily computed using the result 
(\ref{evolvedstate})
\be
L(t) \equiv |\langle \Psi(0)| \Psi(t) \rangle |^{2} 
= \left\vert \sum_{n=0}^{N} {N \choose n}
\cos^{2(N-n)}(\tfrac{\theta}{2})  \sin^{2n}(\tfrac{\theta}{2})  
\sum_{l =0}^{2s}
c_{0,l}^{(s)}(n)\,
e^{-i t \omega_{l}^{(s)}(n) } \right\vert^{2}\,.
\label{Loschmidt}
\ee
The fact that $L(0)=1$ is ensured by the second identity 
in (\ref{identities}).

\begin{figure}[htb]
\centering
{\includegraphics[width=5.5cm]{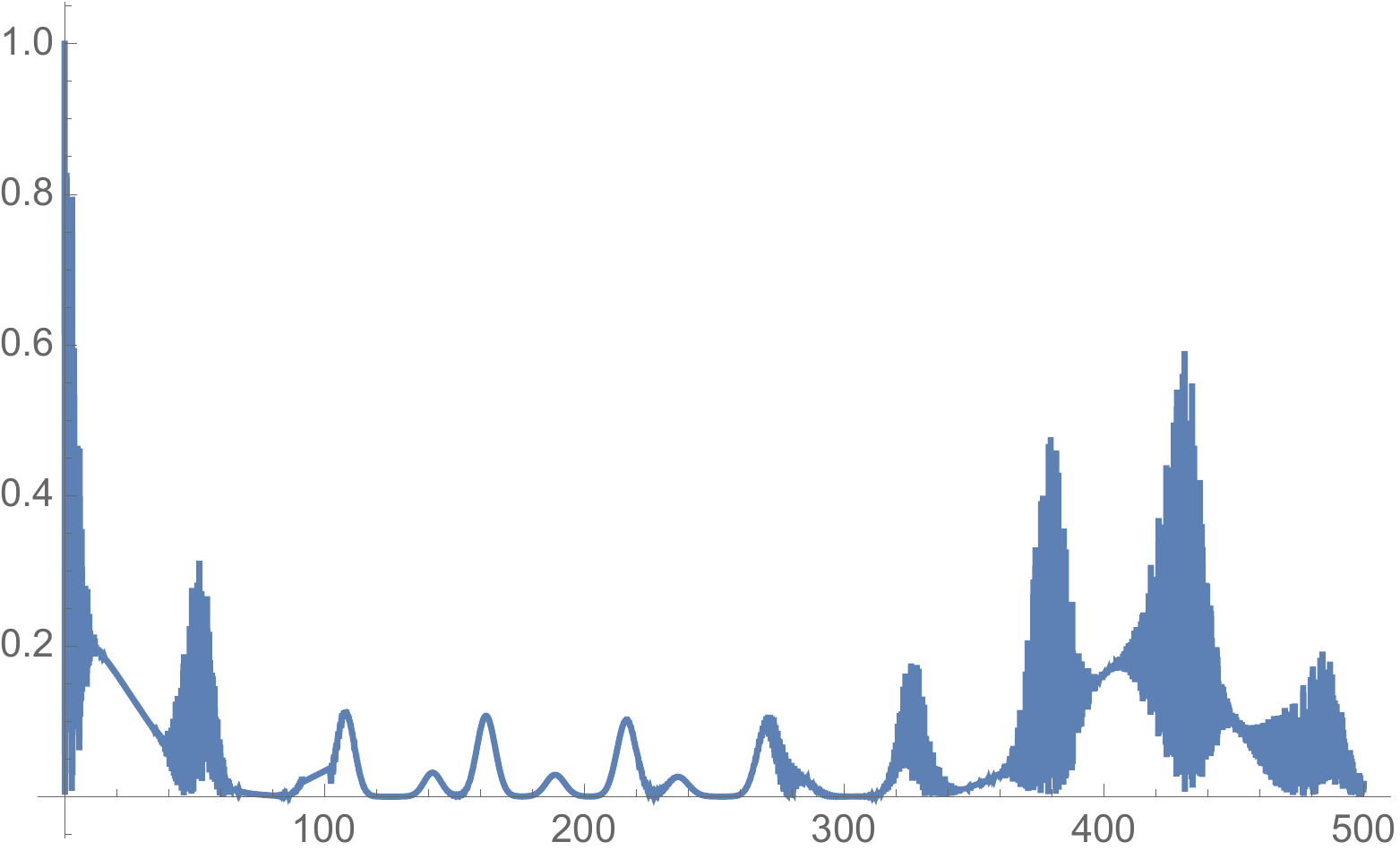}}
\caption{The Loschmidt echo (\ref{Loschmidt})
as a function of time for $N=15, s=1, \theta=0.5 \pi, A=B=1.0$.} 
\label{fig:Loschmidt}
\end{figure}

An example of the Loschmidt echo $L(t)$ for $s=1$ is presented in 
Fig. \ref{fig:Loschmidt}. In contrast with the 
$s=\frac{1}{2}$ case \cite{Guan:2018b}, here $L(t)$ displays rapid 
irregular oscillations. Notice the appearance of points when
$L(t)=0$, at which times the states are completely orthogonal to the 
initial state.

\section{Conclusions}\label{sec:conclusion}

We have presented two approaches for determining the exact spectrum
of the spin-$s$ homogeneous central spin model.  The first approach,
based on properties of $SU(2)$, leads to the reduced eigenvalue problem
(\ref{eigw}).  The second approach, based on integrability, leads to
the Bethe ansatz solution (\ref{EBAhomg}) and (\ref{BAhom}); however,
it remains to be understood whether this solution can give the full spectrum.

We have also performed an exact analysis of the time evolution of a 
spin coherent state, leading to the result (\ref{evolvedstate}).
This allows us to compute the time evolution of various quantities of physical 
interest, including the entanglement entropy (\ref{entropy}), spin 
polarization (\ref{spinpolarization}) and Loschmidt echo 
(\ref{Loschmidt}). For $s>\frac{1}{2}$, we have found interesting differences in 
comparison with the $s=\frac{1}{2}$ case \cite{Guan:2018b}.
While we have restricted for simplicity to the case 
that the central spin is initially ``up'', it should be 
possible to generalize this analysis to other cases, such as the 
GHZ \cite{Greenberger:1990}-like state $(|s, s \rangle + |s, -s 
\rangle)/\sqrt{2}$, or states corresponding to entangled states of 
lower spin. We expect that this model can help usher in the era of tunable quantum 
metrology.

\section*{Acknowledgments}
RN was supported by the Chinese Academy of Sciences President's International Fellowship 
Initiative Grant No. 2018VMA0017, and he is grateful for the hospitality extended to 
him at the Wuhan Institute of Physics and Mathematics, where this 
work was performed. This work was supported by
Key NNSFC grant number 11534014 and by MOST grant number 2017YFA0304500.

\appendix

\section{Computation of the coefficients $h^{(s, 
k)}_{j}$}\label{sec:coeffs}

Here we compute the important coefficients $h^{(s, k)}_{j}$ appearing 
in (\ref{Hk}). Our strategy is to determine these coefficients by means of  
recurrence relations. Acting on both sides of (\ref{Hk}) with the 
Hamiltonian $H^{(s)}$, we 
obtain
\begin{align}
(H^{(s)})^{k+1}\left(|s, s \rangle \otimes |n\rangle \right) &= 
\sum_{j=0}^{2s} h^{(s,k)}_{j} 
H^{(s)} \left(|s,s-j \rangle \otimes |n-j\rangle \right) \non\\
&= \sum_{j=0}^{2s} h^{(s, k+1)}_{j} 
|s,s-j \rangle \otimes |n-j\rangle \,.
\end{align}
Writing the Hamiltonian in terms of spin raising and lowering 
operators (\ref{spinShomogH3}) and making use of 
(\ref{raisinglowering}), we arrive at a system of recurrence 
relations
\be
h^{(s,k+1)}_{j}  = \alpha_{j}^{(s)}\, h^{(s,k)}_{j} + 
\beta_{j}^{(s)}\, h^{(s,k)}_{j+1} + 
\beta_{j-1}^{(s)}\, h^{(s,k)}_{j-1}\,, \qquad h^{(s,0)}_{j}  = \delta_{j,0} \,, 
\qquad j = 0, 1, \ldots, 2s\,,
\label{recurrence}
\ee
whose coefficients are given by
\begin{align}
   \alpha_{j}^{(s)} &=B (s-j) + 2 A(s-j)(\frac{N}{2} + j-n) \,, \non\\ 
   \beta_{j}^{(s)} &= A \sqrt{(j+1)(2s-j)(n-j)(N+j+1-n)} \,.
   \label{alphabeta}
\end{align}   
In order to solve these recurrence relations, we 
define the generating functions
\be
h_{j}^{(s)}(z) = \sum_{k=0}^{\infty} h^{(s,k)}_{j} z^{k} \,.
\label{genfuncs}
\ee
By virtue of (\ref{recurrence}), these generating functions satisfy a system of linear relations
\be
(\alpha_{j}^{(s)} - \frac{1}{z})\, h_{j}^{(s)}(z) + \beta_{j}^{(s)}\, h_{j+1}^{(s)}(z) + 
\beta_{j-1}^{(s)}\, h_{j-1}^{(s)}(z) = - \frac{1}{z} \delta_{j,0} \,, 
\qquad j = 0, 1, \ldots, 2s \,,
\label{genfuncrltns}
\ee
which we can also write in matrix form
\be
M^{(s)}(z)\, h^{(s)}(z) =  C^{(s)}(z)\,,
\label{matform}
\ee
where we have introduced the column vectors $h^{(s)}(z)  = 
(h_{0}^{(s)}(z), \ldots, h_{2s}^{(s)}(z))^{T}$ and $C^{(s)}(z) = (- 
\frac{1}{z}, 0, \ldots , 0)^{T}$, and $M^{(s)}(z)$ is the tridiagonal 
matrix\footnote{We suppress here the superscripts ${}^{(s)}$ on the 
$\alpha$'s and $\beta$'s.}
\be
M^{(s)}(z) = \left(\begin{array}{ccccccc}
\alpha_{0}-\frac{1}{z} & \beta_{0} & 0 & 0  &\ldots & 0 & 0 \\
\beta_{0} & \alpha_{1}- \frac{1}{z} & \beta_{1} & 0 & \ldots & 0 & 0 \\
0 & \beta_{1} & \alpha_{2}- \frac{1}{z} & \beta_{2} & \ldots & 0 & 0 \\
\vdots & \vdots & \vdots & \vdots &  & \vdots & \vdots \\
0 & 0 & 0 & 0 & \ldots &  \beta_{2s-1} & \alpha_{2s}-\frac{1}{z}
\end{array}
\right)_{(2s+1) \times (2s+1)} \,.
\label{Mmat}
\ee

The solutions of the relations (\ref{genfuncrltns}) are given by the 
following rational functions of $z$
\be
 h_{j}^{(s)}(z)  = \frac{\sum_{l=0}^{2s} n_{l}^{(s, j)} 
 z^{l}}{\sum_{l=0}^{2s+1} d_{l}^{(s)} z^{l}} \,, \qquad j = 0, 1, \ldots, 2s 
 \,.
\label{hjzsltn}
\ee
The coefficients in the denominator $d_{l}^{(s)}$ can be read 
off from an expansion of the determinant of the matrix $M^{(s)}(z)$ 
(\ref{Mmat}) in inverse powers of $z$,
\be
\det \left( M^{(s)}(z) \right) =\sum_{l=0}^{2s+1} \frac{ (-1)^{2s+1} }{z^{2s+1-l}} 
d^{(s)}_{l} \,, \qquad d_{0}^{(s)} = 1 \,.
\label{denomcoeffs}
\ee
Note that $d^{(s)}_{l}$ is independent of the value of $j$ in 
(\ref{hjzsltn}). The coefficients in the numerator $n_{l}^{(s, j)}$, 
which do depend on the value of $j$, can be read off from a similar
expansion of the {\em minors} of the first row of the matrix 
$M^{(s)}(z)$,
\be
{\rm minor}_{(1,1+j)} \left( M^{(s)}(z) \right) =\sum_{l=0}^{2s} 
\frac{ (-1)^{2s+j} }{z^{2s-l}} 
n^{(s,j)}_{l} \,, \qquad n^{(s,j)}_{0} = \delta_{j,0}\,, \qquad j = 0, 1, \ldots, 2s
\label{numcoeffs}
\,.
\ee
Clearly, both  $d^{(s)}_{l}$ and $n_{l}^{(s, j)}$ are expressed in 
terms of the $\alpha$'s and $\beta$'s (\ref{alphabeta}).

Performing a partial fraction decomposition of the solutions 
$h_{j}^{(s)}(z)$ (\ref{hjzsltn}), we obtain
\be
h_{j}^{(s)}(z) = \sum_{l=0}^{2s}\frac{c_{j,l}^{(s)}}{1-\omega_{l}^{(s)} z}\,,
\label{hpf}
\ee
where $\omega_{l}^{(s)}$ is the $(l+1)^{th}$ root of the following polynomial 
equation of degree $2s+1$
\be
\sum_{i=0}^{2s+1} d_{i}^{(s)}\, z^{2s+1-i} = 0 \,.
\label{gpoly}
\ee
Moreover, we obtain formulas for
$c_{j,l}^{(s)}$ by evaluating the residues of $h_{j}^{(s)}(z)$ 
(\ref{hjzsltn}) at $z=1/\omega_{l}^{(s)}$, 
\be
c_{j,l}^{(s)} = \frac{\sum_{i=0}^{2s} n_{i}^{(s, j)} 
\left(\omega_{l}^{(s)}\right)^{2s-i} }{\prod_{i=1, i\ne 
l}^{2s+1}(\omega_{l}^{(s)} - \omega_{i}^{(s)})} \,.
\label{cjl}
\ee

In view of the definition (\ref{genfuncs}) of the generating 
functions, the sought-after 
coefficients $h_{j}^{(s,k)}$ can be obtained from
\be
h_{j}^{(s,k)} = \frac{1}{k!} \frac{d^{k}}{dz^{k}} h_{j}^{(s)}(z) 
\Big\vert_{z=0} \,.
\ee
Performing this computation using the solution (\ref{hpf}), we 
conclude that 
\be
h_{j}^{(s,k)} = \sum_{l=0}^{2s} c_{j,l}^{(s)}\, 
\left(\omega_{l}^{(s)}\right)^{k} \,, \qquad j = 0, 1, \ldots, 2s 
\,.
\label{hjsltn}
\ee 
As a consistency check, note that substituting the result 
(\ref{hjsltn}) back into (\ref{genfuncs}) and interchanging the order of 
summations, one obtains
\be
h_{j}^{(s)}(z) = \sum_{l=0}^{2s} c_{j,l}^{(s)} \left[ 
\sum_{k=0}^{\infty} \left(\omega_{l}^{(s)} z \right)^{k} \right] \,.
\label{geometric}
\ee
Summing the geometric series in (\ref{geometric}), one recovers the result (\ref{hpf}).

In short, we have the following recipe for computing the coefficients 
$h_{j}^{(s,k)}$ appearing in (\ref{Hk}):
\begin{enumerate}
    \item Construct the matrix $M^{(s)}(z)$  (\ref{Mmat}), where the 
    $\alpha$'s and $\beta$'s are defined in (\ref{alphabeta}).
    \item Read off $d_{l}^{(s)}$ and $n_{l}^{(s, j)}$ using
    (\ref{denomcoeffs}) and  (\ref{numcoeffs}), respectively.
    \item Solve the polynomial equation (\ref{gpoly}) to obtain 
    $\omega_{l}^{(s)}$.
    \item Obtain $c_{j,l}^{(s)}$ using (\ref{cjl}).
    \item Obtain $h_{j}^{(s,k)}$ using (\ref{hjsltn}).
    \end{enumerate}
It is straightforward to implement this recipe numerically on a 
computer. 

Note that $\omega_{l}^{(s)}$ and $c_{j,l}^{(s)}$ also depend on 
$n$, $N$, $A$, and $B$ (through the $\alpha$'s and $\beta$'s), but we 
have not explicitly displayed these dependences here in order to lighten the 
notation. However, we {\em do} display the dependence on $n$ in the 
corresponding formula in the body of the paper (\ref{hjsltnmain}), 
due to the presence there of a summation over $n$.

\section{Identities for $c_{j,l}^{(s)}$}\label{sec:identities}

We obtain here some useful identities for the coefficients 
$c_{j,l}^{(s)}$ appearing in $h_{j}^{(s,k)} $ (\ref{hjsltn}).
We begin by evaluating the generating functions $h_{j}^{(s)}(z)$ at 
$z=0$ in two different ways: using (\ref{hjzsltn}) we obtain
\be
h_{j}^{(s)}(0) = \frac{n_{0}^{(s, j)}}{d_{0}^{(s)}} = \delta_{j,0} \,,
\label{id1a}
\ee
while (\ref{hpf}) gives
\be
h_{j}^{(s)}(0) = \sum_{l=0}^{2s} c_{j,l}^{(s)}\,.
\label{id1b}
\ee
We conclude from (\ref{id1a}) and (\ref{id1b}) that
\be
\sum_{l=0}^{2s} c_{j,l}^{(s)} = \delta_{j,0}  \,.
\label{id1}
\ee
Multiplying both sides of (\ref{id1}) by $c_{j,l'}^{(s)}$ and summing 
over $j$, we obtain
\be
\sum_{j=0}^{2s} \sum_{l=0}^{2s} c_{j,l}^{(s)}\, c_{j,l'}^{(s)} =  c_{0,l'}^{(s)}  \,.
\label{id2}
\ee

Let us now evaluate the expression $\sum_{j=0}^{2s} h_{j}^{(s)}(z)\, 
h_{j}^{(s)}(z')$ in two different ways. Using  (\ref{hpf})  we obtain
\be
\sum_{j=0}^{2s} h_{j}^{(s)}(z)\, h_{j}^{(s)}(z') = \sum_{j=0}^{2s} 
\sum_{l, l'=0}^{2s}\frac{c_{j,l}^{(s)}\, c_{j,l'}^{(s)}}{(1-\omega_{l}^{(s)} 
z)(1-\omega_{l'}^{(s)} z')} \,.
\label{id3}
\ee 
On the other hand, we know from (\ref{matform}) that $h^{(s)}(z) = 
(M^{(s)}(z))^{-1}\, C^{(s)}(z)$. Therefore,
\be
\sum_{j=0}^{2s} h_{j}^{(s)}(z)\, h_{j}^{(s)}(z') = \frac{1}{z z'}\left( M(z) M(z') 
\right)^{-1}_{00} \,.
\label{other}
\ee
Since $\det(M(z))$ does not vanish at $z=\omega_{l}^{(s)}$ (it 
vanishes instead at $z=1/\omega_{l}^{(s)}$), and similarly for 
$\det(M(z'))$, we see from (\ref{other}) that $\sum_{j=0}^{2s} h_{j}^{(s)}(z)\, 
h_{j}^{(s)}(z')$ is regular at $z=\omega_{l}^{(s)}\,, 
z'=\omega_{l'}^{(s)}$. Evaluating the residues of both sides of (\ref{id3}) 
at $z=\omega_{l}^{(s)}\,, z'=\omega_{l'}^{(s)}$ with $l \ne 
l'$, we obtain
\be
\sum_{j=0}^{2s} c_{j,l}^{(s)}\, c_{j,l'}^{(s)} =  0 \,, \qquad l \ne 
l'  \,.
\label{id4}
\ee
Combining this result with (\ref{id2}), we conclude that
\be
\sum_{j=0}^{2s} c_{j,l}^{(s)}\, c_{j,l'}^{(s)} = c_{0,l}^{(s)} \, 
\delta_{l,l'}  \,.
\label{id5}
\ee
The main results of this appendix are the identities (\ref{id1}) and (\ref{id5}).

% \newpage
% \clearpage

% \bibliographystyle{utphys}
% \bibliography{refs}

\providecommand{\href}[2]{#2}\begingroup\raggedright\endgroup

\end{document}